\documentclass{vietnam}

\def\NIMA{{\em Nucl. Instrum. Methods} A}

\def\PLB{{\em Phys. Lett.}  B}
\def\PRL{{\em Phys. Rev. Lett.}}
\def\PRD{{\em Phys. Rev.} D}

\def\etal{{\em et al.}}
\def\icrc77{{\em Proc. of the 15th ICRC}}
\def\ICRC13{{\em Proc. of the 33rd ICRC}}

\def\be{\begin{equation}}
\def\ee{\end{equation}}
\def\bea{\begin{eqnarray}}
\def\eea{\end{eqnarray}}

\def\xmax{$X_\mathrm{max}$}
\def\m_xmax{$\langle X_\mathrm{max}\rangle$}
\def\rms_xmax{$\sigma(X_\mathrm{max})$}

\newcommand{\Photo}{}

\begin{document}
\vspace*{4cm}
\title{Measurement of the chemical composition of the ultra-high-energy cosmic rays with the Pierre Auger Observatory}

\author{Matthias Plum \footnotemark[1]$^{,*}$ \footnotetext{$^{*}$E-mail: \url{plum@physik.rwth-aachen.de}} for the Pierre Auger Collaboration \footnotemark[2]}
\address{\footnotemark[1]RWTH Aachen University, Physics Institute III A, Aachen, Germany\\
\footnotemark[2]Observatorio Pierre Auger, Av San Martin Norte 304, 5613 Malarg\"ue, Argentina\\
(Full author list: \url{http://www.auger.org/archive/authors_2013_06.html})
}

\maketitle\abstracts{
The Pierre Auger Observatory infers the chemical composition of ultra-high-energy cosmic rays through two independent detection techniques. The Fluorescence Detector (FD) measures the longitudinal profile of high energy air showers and can determine the depth of the shower maximum \xmax, which is sensitive to the chemical composition of the primary cosmic rays. Additionally, measurements by the Surface Detector (SD) provide independent experimental observables based on the muonic shower component to analyze the chemical composition.\\
We present the results for the \xmax\ distributions and the mass composition results measured by the FD and the SD for the energies $E \geq 10^{18}$\,eV. The data will be compared with the expectations for proton and iron primaries according to different hadronic interaction models.}

\section{Introduction}
The chemical composition of ultra-high-energy cosmic rays (UHECRs) with energies E\,$\geq 10^{18}$\,eV can only be studied indirectly by measuring the secondary particles produced in extensive air showers. The determination of the composition is important to understand the origin of UHECRs. In this work, we show some of the latest measurements of composition related observables of the Pierre Auger Observatory in comparison with different interaction models.

\subsection{Pierre Auger Observatory}
The Pierre Auger Observatory is the world largest detector for UHECRs with an aperture of 3000\,km$^2$ covered by the Surface Detector array \cite{SD} (SD) of 1660 water-Cherenkov stations and overlooked by the Fluorescence Detector \cite{FD} (FD) with 27 telescopes at the borders of the array. This hybrid design with 2 independent detector components allows one to study simultaneously the longitudinal shower profile in the atmosphere and the lateral distribution of shower particles on the ground with high accuracy. Due to the low fluorescence light flux of extensive air showers the FD can only take data during clear, moonless nights, which corresponds to a duty cycle of 13\,\%, whereas the SD can measure with a duty cycle of nearly 100\,\%.
\section{Studies of the Chemical Composition}
The Pierre Auger Observatory measures the primary particle composition with both detector components, the FD and the SD. In this work we present two different analyses.
\subsection{Fluorescence Detector Analysis}
One way to determine the mass of the primary cosmic ray is to study the longitudinal shower profile of the electromagnetic component in the atmosphere with the FD and to measure the depth of the shower maximum \xmax, which is the position of the maximum of energy deposition per atmospheric slant depth of an extensive air shower \cite{PRL,desouza}. Lighter primaries penetrate the atmosphere deeper than heavier primaries. Also, due to the larger number of nucleons and the larger cross section, the event-by-event fluctuations of \xmax\ should be smaller for heavier nuclei. So the first two moments of the \xmax\ distribution, which are the mean \m_xmax and standard deviation \rms_xmax are used to study the mass composition in the energy range above $10^{17.8}$\,eV.
\begin{figure}
\centerline{\includegraphics[width=0.7\linewidth]{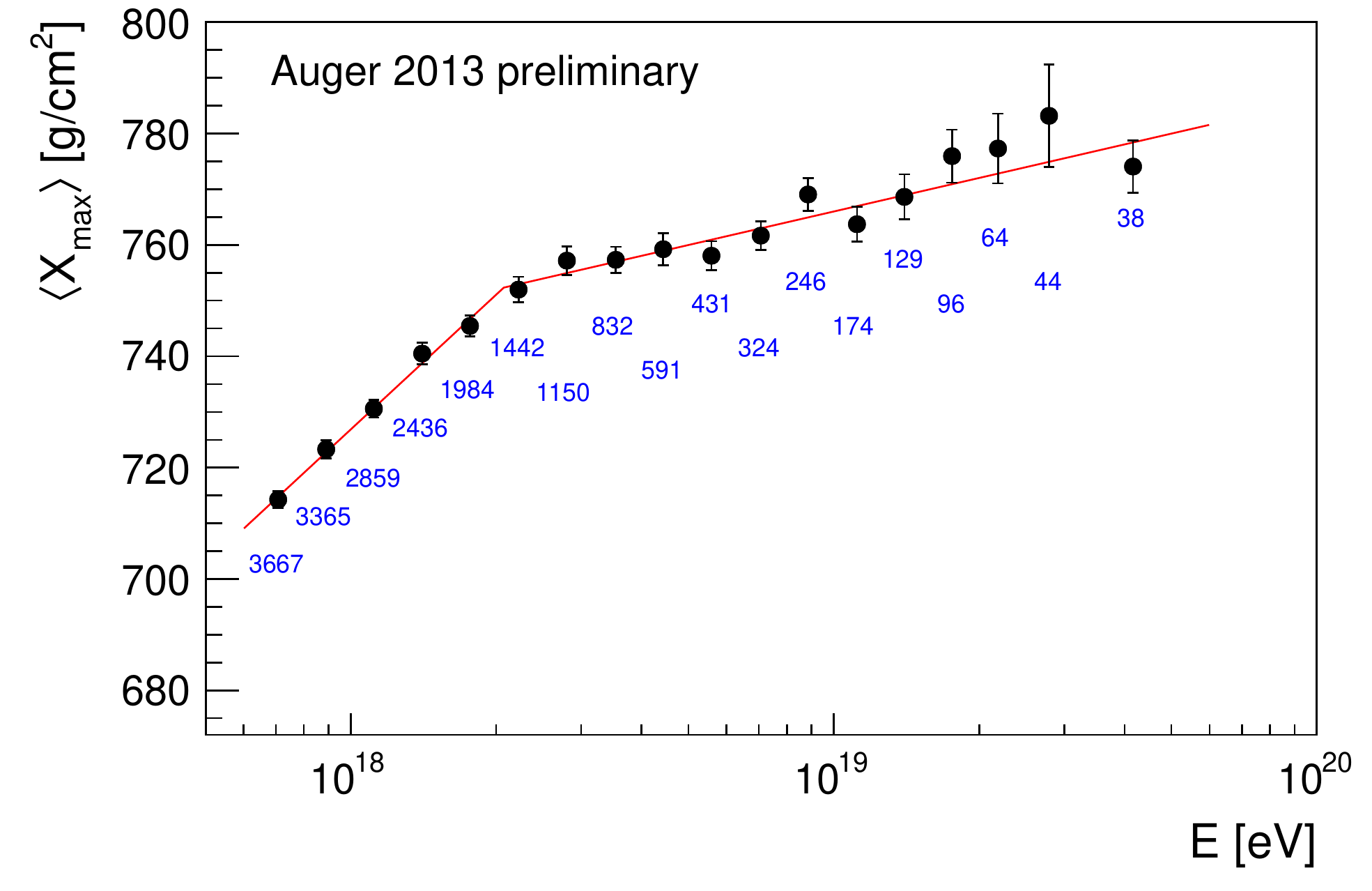}}
\caption{Development of the mean depth of the shower maximum \m_xmax as a function of the energy, with a broken line fit of the elongation rate \protect\cite{antoine}.}
\label{fig:icrc_linefit}
\end{figure}
In figure \ref{fig:icrc_linefit} the average \m_xmax is compared with energy \cite{antoine}. It shows that the elongation rate $D$, which is defined as
\begin{equation}
D = \frac{d \langle X_\mathrm{max}\rangle}{d\,\log\,E} ,
\label{eq:elong_rate}
\end{equation}
cannot be described by a single line fit ($\chi^2/ndf = 128.1/16$), but requires a broken line fit ($\chi^2/ndf = 10.3/14$). This suggests a change in the composition of UHECRs with increasing energy.
\begin{figure}
\centerline{\includegraphics[width=1.0\linewidth]{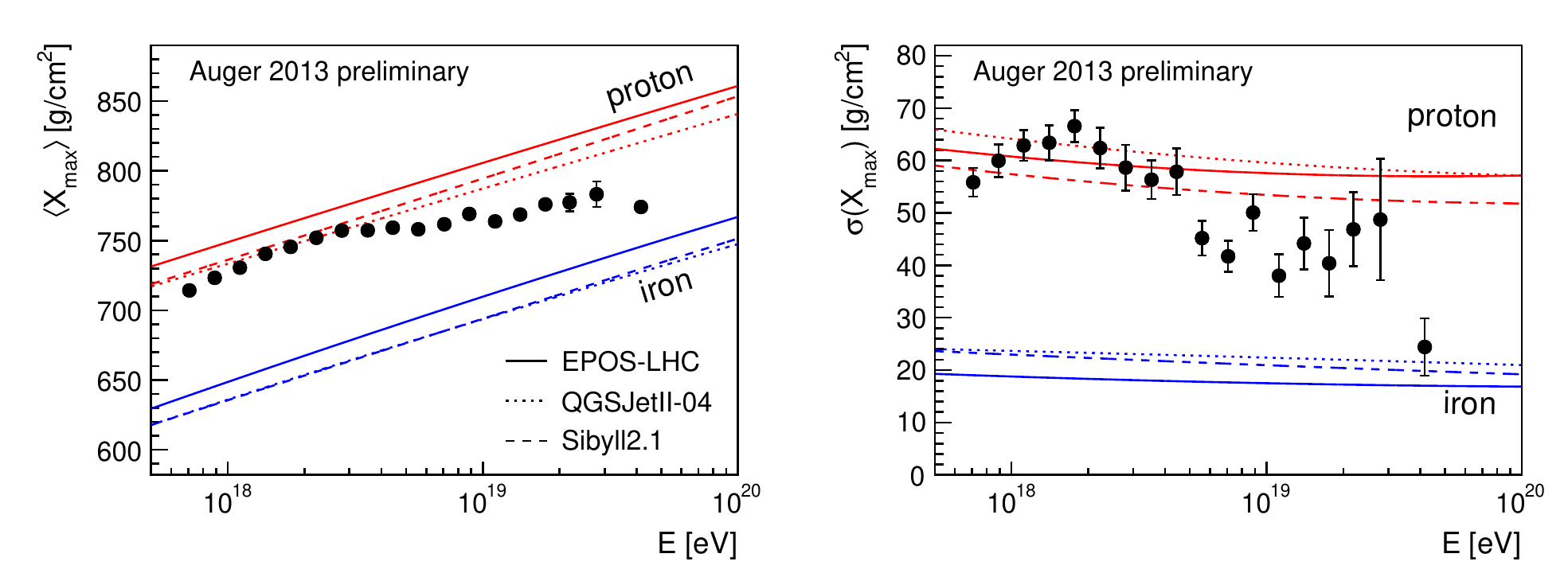}}
\caption{Energy evolution \protect\cite{antoine} of the average of the shower maximum \m_xmax (left) and event-by-event fluctuations \rms_xmax (right) compared to the high-energy interaction models Sibyll2.1 \protect\cite{sibyll}, EPOS-LHC \protect\cite{epos} and QGSJetII-04 \protect\cite{qgsjet}.}
\label{fig:icrc_xmax_model}
\end{figure}
In figure \ref{fig:icrc_xmax_model} the energy evolution \cite{antoine} of the average of the shower maximum \m_xmax and event-by-event fluctuations \rms_xmax are shown in comparison with the recent interaction models for extensive air showers Sibyll2.1 \cite{sibyll}, EPOS-LHC \cite{epos} and QGSJetII-04 \cite{qgsjet}. This comparison suggests that with increasing energy there is a transition from a lighter to a heavier mass composition \cite{ahn}. However, this interpretation is strongly dependent on the interaction models.
\subsection{Surface Detector Analysis}
\begin{figure}
\centerline{\includegraphics[width=0.65\linewidth]{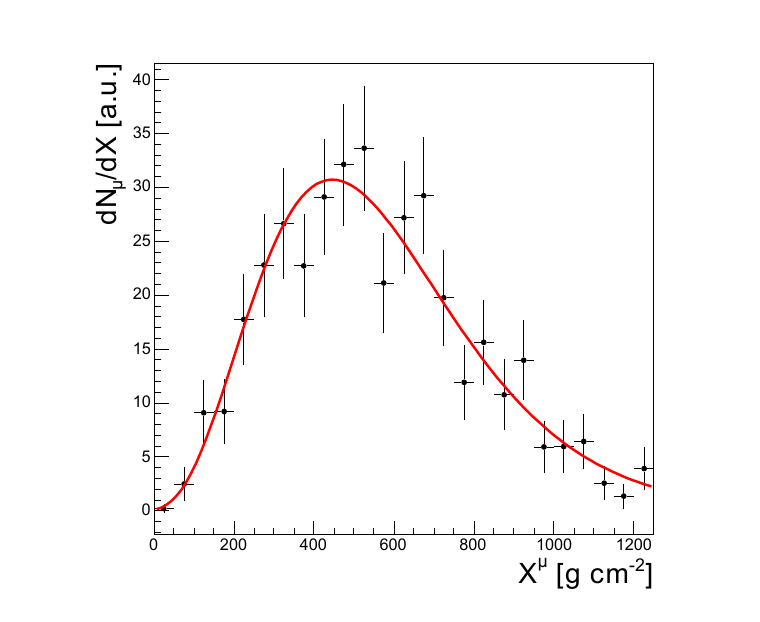}}
\caption{Reconstructed muon production depth (MPD) \protect\cite{garcia} of an extensive air shower with a zenith angle $\theta = (59.06\, \pm\, 0.08)^\circ$ and an energy $E = (92\, \pm\, 3)$\,EeV with a Gaisser-Hillas function fit \protect\cite{gaisser}.}
\label{fig:icrc_xmumax_long}
\end{figure}

\begin{figure}
\centerline{\includegraphics[width=0.7\linewidth]{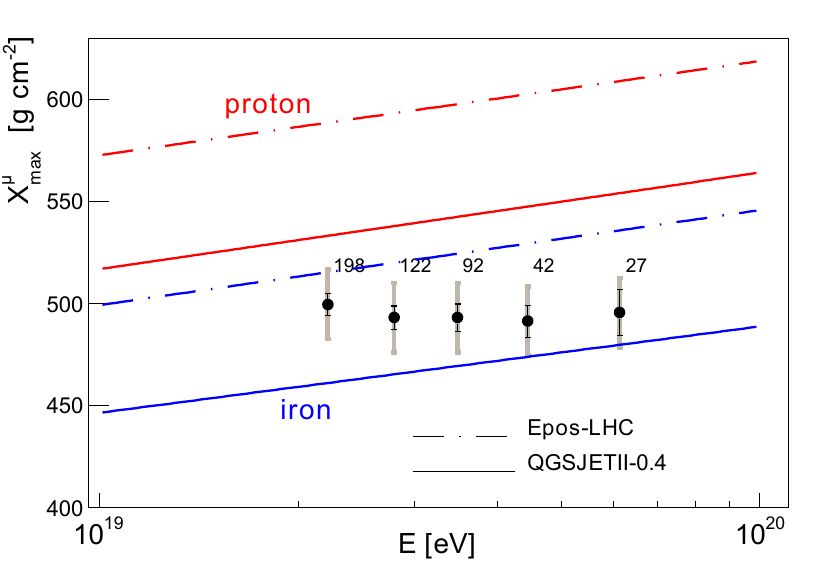}}
\caption{Energy evolution of $\langle X^{\mu}_\mathrm{max}\rangle$ compared with interaction models EPOS-LHC \protect\cite{epos} and QGSJetII-04 \protect\cite{qgsjet}. The gray rectangles represent the systematic uncertainties \protect\cite{garcia}.}
\label{fig:icrc_xmumax}
\end{figure}

It is possible to study the primary composition of UHECRs by reconstruction of the muon production depth (MDP) along the shower axis \cite{garcia}. Due to the fact that muons come from pion and kaon decays, the MPD allows to study the hadronic component of the shower. This can be inferred from the timing information of muons in the SD stations, and a geometrical back-projection to the shower axis. In this way, SD stations can be used to reconstruct the longitudinal shower profile. However, signals measured in the SD are a mix of the electromagnetic and the muonic component of the air shower. To reduce the electromagnetic contamination, only air showers in the zenith angle band $55^\circ < \theta< 65^\circ$ are included and to reduce distortions in the reconstructed depths, only SD stations which are in the range of 1700\,m to 4000\,m from the shower core are taken. By means of the longitudinal distribution of the MPD, the $X^{\mu}_{max}$, which is the maximum of produced muons per atmospheric slant depth, can be reconstructed by a fit with a Gaisser-Hillas function \cite{gaisser}, as shown in figure \ref{fig:icrc_xmumax_long}. 
Similar to the \m_xmax study in the previous section, the mean $\langle X^{\mu}_\mathrm{max}\rangle$ based on Monte-Carlo studies is smaller for heavier and larger for lighter primary cosmic rays. In figure \ref{fig:icrc_xmumax} the data of this study are shown for showers with an energy $E \geq 20$\,EeV. The interpretation of these data in terms of chemical composition is very challenging due to the considerable differences in the muon production number \cite{kegl,valino} and elongation rate based on the interaction models as shown in figure \ref{fig:icrc_xmumax}. This study can be used to constrain the interaction models and improve the understanding of hadronic interactions.
The slope of the elongation rate suggests a change in the composition of UHECRs. However, this study currently suffers from limited statistics and the resolution on the MPD. Therefore, the measurements currently cannot rule out a constant composition. 
\section{Future prospects}
The sensitivity of measurements to the chemical composition of UHECRs will increase considerably in the next few years through increased statistics and by improved analyses. Additionally, to improve the sensitivity to the composition, the Pierre Auger Collaboration plans to enhance the electronics of the SD stations to improve the MPD resolution, and to add a dedicated muon detection system to the ground array to perform an event-by-event measurement of the muonic longitudinal shower profile with with large statistics \cite{future}.


\section*{References}
\begin{thebibliography}{99}
\bibitem{SD} J. Abraham \etal, \NIMA, \textbf{523}, 50 (2004)
\bibitem{FD} J. Abraham \etal, \NIMA, \textbf{620}, 227 (2010)
\bibitem{PRL} The Pierre Auger Coll., \PRL, \textbf{104}, 091101 (2010)
\bibitem{desouza} V. de Souza for the Pierre Auger Coll., \ICRC13, arXiv:1307.5059 [astro-ph.HE] (2013)
\bibitem{antoine} A. Letessier-Selvon for the Pierre Auger Coll., \ICRC13, arXiv:1310.4620 [astro-ph.HE] (2013)
\bibitem{sibyll} R. Fletcher \etal, \PRD, \textbf{50}, 5710 (1994)
\bibitem{epos} T. Pierog \etal, arXiv:1306.0121 [hep-ph] (2013)
\bibitem{qgsjet} S. Ostapchenko, \PLB, \textbf{703}, 588 (2011)
\bibitem{ahn} E. Ahn for the Pierre Auger Coll., \ICRC13, arXiv:1307.5059 [astro-ph.HE] (2013)
\bibitem{garcia} D. Garcia-Gamez for the Pierre Auger Coll., \ICRC13, arXiv:1307.5059 [astro-ph.HE] (2013)
\bibitem{gaisser} T. Gaisser and A. Hillas, \icrc77,  Vol. 8, p. 353. (1977)
\bibitem{kegl} B. Kegl for the Pierre Auger Coll., \ICRC13, arXiv:1307.5059 [astro-ph.HE] (2013)
\bibitem{valino} I. Vali\~{n}o for the Pierre Auger Coll., \ICRC13, arXiv:1307.5059 [astro-ph.HE] (2013)
\bibitem{future} The Pierre Auger Coll., arXiv:1307.0226 [astro-ph.IM] (2013)
\end{thebibliography}
\end{document}